\documentclass[doublecol]{epl2}
\usepackage{amsmath}
\usepackage[normalem]{ulem}
\newcommand{\revis}[1]{\textcolor{black}{#1}}
\begin{document}
\title{Statistical mechanics of aggregation and crystallization for\\ semiflexible polymers}
\author{Christoph Junghans \inst{1,2}\thanks{E-mail: \email{junghans@mpip-mainz.mpg.de}}\ \and 
Michael Bachmann \inst{1,3}\thanks{E-mail: \email{m.bachmann@fz-juelich.de}} \and 
Wolfhard Janke \inst{1}\thanks{E-mail: \email{Wolfhard.Janke@itp.uni-leipzig.de}}}
\shortauthor{C.\ Junghans, M.\ Bachmann and W.\ Janke}
\institute{
  \inst{1} Institut f\"ur Theoretische Physik and Centre for Theoretical Sciences (NTZ), 
Universit\"at Leipzig -- Postfach 100\,920, D-04009 Leipzig, Germany, EU\\
  \inst{2} Max-Planck-Institut f\"ur Polymerforschung -- Ackermannweg 10, D-55128 Mainz, Germany, EU\\
  \inst{3} Institut f\"ur Festk\"orperforschung, Theorie II, Forschungszentrum J\"ulich -- D-52425 
J\"ulich, Germany, EU
}
\abstract{
By means of multicanonical computer simulations, we investigate thermodynamic properties 
of the aggregation of interacting semiflexible polymers. We analyze a mesoscopic bead-stick model, where
nonbonded monomers interact via \revis{Lennard-Jones} forces. Aggregation turns out to 
be a process, in which the constituents experience strong structural fluctuations, 
similar to peptides in coupled folding-binding cluster formation processes. In contrast
to a recently studied related proteinlike hydrophobic-polar heteropolymer model, 
aggregation and crystallization are separate processes \revis{for a homopolymer
with the same small bending rigidity. Rather stiff semiflexible polymers form a 
liquid-crystal-like phase, as expected. In} analogy to the heteropolymer
study, we find that the first-order-like aggregation transition of the 
complexes is accompanied by strong system-size dependent hierarchical surface
effects. In consequence, the polymer aggregation is a phase-separation process
with entropy reduction.
}
\maketitle
Cluster formation and crystallization of polymers are processes which are 
interesting for technological applications, e.g., \revis{for the design of} new materials
with certain mechanical properties or
nanoelectronic organic devices and polymeric solar cells. 
From a biophysical point of view, the understanding of peptide oligomerization, but also the
(de)fragmentation
in semiflexible biopolymer systems like actin networks
is of substantial relevance. This requires
a systematic analysis of \revis{the basic properties of the} polymeric 
cluster formation processes,
in particular, for small polymer complexes on the nanoscale, where 
surface effects are 
competing noticeably with structure-formation processes in the interior of the 
aggregate. 

\revis{A further motivation for investigating the aggregation transition of 
semiflexible {\em homo\/}polymer chains derives from
the intriguing results of our recent study of a similar aggregation 
process for peptides~\cite{jbj1,jbj2}, which were modeled as {\em hetero\/}polymers
with a sequence of two types of monomers, hydrophobic ($A$) and hydrophilic ones ($B$). By 
specializing the previously employed heteropolymer model to 
the apparently simpler homopolymer case, we aim by comparison at isolating
those properties which are mainly driven by the sequence properties of 
heteropolymers. In fact, while in both cases the aggregation transition is a
phase-separation process, we will show below that for homopolymers the 
aggregation and crystallization (if any) are separate conformational
transitions -- unlike our study of heteropolymer aggregates where they 
were found to coincide~\cite{jbj1,jbj2}. The physical origin causing these
differences will be explained within the microcanonical
formalism~\cite{gross1,thirring1}, which proves~\cite{jbj1,jbj2} to be particularly
suitable for this type of problem.}

\revis{We thus consider the
same  model as in~\cite{jbj1,jbj2}, but here we assume
that} all monomers $i_\mu=1,\ldots,N^{(\mu)}$ of the $\mu$th
chain ($\mu=1,\ldots,M$) at positions 
$\textbf{x}_{i_\mu}$ are hydrophobic ($A$).
The bonds between adjacent monomers are taken to be rigid (bead-stick model) and
pairwise interactions among nonbonded monomers 
are modeled by a Lennard-Jones potential 
\begin{equation}
V_{\rm LJ}(r_{i_\mu j_\nu})=4[r_{i_\mu j_\nu}^{-12}-r_{i_\mu j_\nu}^{-6}]\, ,
\end{equation}
where $r_{i_\mu j_\nu}=|\textbf{x}_{i_\mu}-\textbf{x}_{j_\nu}|$ is the distance 
between monomers $i_\mu$ and $j_\nu$ of the $\mu$th and $\nu$th chain, respectively. 
\revis{In the peptide model, the coefficient of the van der Waals contribution 
$\propto r^{-6}$ would depend on the type of monomers involved \cite{jbj1,jbj2}.}
Intra-chain ($\mu=\nu$) and inter-chain ($\mu\neq\nu$) contacts
are not distinguished energetically. The semiflexibility of a chain is described 
by the bending energy 
\begin{equation}
E^{(\mu)}_\text{bend}=\kappa\sum_{i_\mu}(1-\cos\vartheta_{i_\mu})\, ,
\end{equation}
where $0\le\vartheta_{i_\mu}\le\pi$ is the bending angle formed by the monomers 
$i_\mu$, $i_\mu+1$, and $i_\mu+2$. For the
comparison with our recent heteropolymer aggregation studies~\cite{jbj1,jbj2},
we choose \revis{in most simulations} a bending rigidity $\kappa=0.25$,
 which is at the rather floppy end 
of semiflexibility.   
Thus, the single-chain energy reads 
\begin{equation}
E^{(\mu)}=E^{(\mu)}_\text{bend}+\sum_{j_\mu>i_\mu+1} V_{\rm LJ}(r_{i_\mu j_\mu})
\end{equation}
and the total energy of the polymer system is given by
\begin{equation}
E=\sum_\mu E^{(\mu)}+\sum_{\mu<\nu}\sum_{i_\mu,j_\nu} V_{\rm LJ}(r_{i_\mu j_\nu})\, .
\end{equation}
All chains are assumed to have the \revis{same degree of polymerisation, i.e., the} 
same number of monomers, $N^{(\mu)}=N$, $\mu = 1,\dots,M$. 

We have performed multicanonical computer simulations~\cite{muca} for different system 
sizes. The multicanonical weights were calculated recursively~\cite{muca2} in 
50 iterations with 
$5\times10^6\times M^2$ single updates (spherical pivot rotations~\cite{baj1}, 
semilocal crankshaft moves~\cite{jbj2}) each.
The production run with fixed multicanonical weights included $5\times 10^8\times M^2$ single 
updates. 
The generic result obtained from these simulations is a precise estimate for the density
of states $g(E)$ or microcanonical entropy $S(E)=k_B \ln g(E)$, \revis{which will be 
central in the second part of our data analysis.} 

\begin{figure}
\onefigure[width=0.49\textwidth]{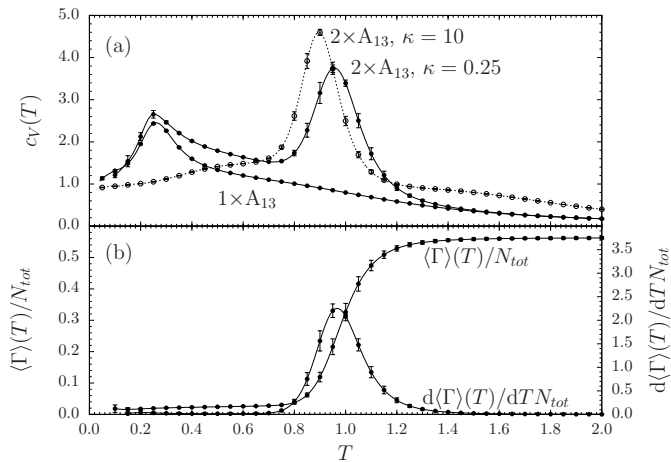}
\caption{
\label{fig:2xA13}
(a) Specific heat per monomer as a function of temperature for a single semiflexible homopolymer 
with 13 monomers ($1\times A_{13}$, $\kappa=0.25$) and for systems of two such chains ($2\times A_{13}$)
with different bending rigidities.
(b) Canonical expectation value $\langle\Gamma\rangle$ 
and fluctuation $d\langle\Gamma\rangle/dT$ of the aggregation parameter $\Gamma$, defined
in Eq.~(\ref{eq:gamma}), for the two-chain system with $\kappa=0.25$. 
}
\end{figure}

From a more standard canonical perspective, 
\revis{conformational transitions between structural macrostates are signalized by 
peaks in the temperature-dependent fluctuations of energy, i.e., the 
specific heat per monomer}
\begin{equation}
c_V=(\langle E^2\rangle-\langle E\rangle^2)/N_{\rm tot}k_BT\, ,
\end{equation}
where $N_{\rm tot}=NM$ and $k_B=1$ in our units. \revis{Knowing $g(E)$, this can be 
straightforwardly computed for any temperature $T$.} In Fig.~\ref{fig:2xA13}(a), the
specific-heat curve for a system of two identical semiflexible polymers ($2\times A_{13}$)
is compared with the energetic fluctuations
of a single chain ($1\times A_{13}$). The single
chain exhibits a very weak coil-globule collapse transition (shoulder near 
$T\approx 0.88$), whereas
the crystallization near $T\approx 0.24$ is a pronounced, separate process.
The thermodynamic phase behavior of single semiflexible polymers in solvent
has already been subject of numerous studies, with particular focus on stiffness
and finite chain length effects~\cite{doniachA,ivanovA,stukanA}, where it was
shown that the \revis{globule-solid} transition is more influenced by stiffness 
effects than the coil-globule transition.
\revis{In particular, f}or longer chains, it was found that, depending on the stiffness,
\emph{single} collapsed semiflexible polymers form spherical, ellipsoidal,
and disklike globules, as well as toroids~\cite{stukanA}.
Our first result
for the semiflexible \emph{multiple}-chain system 
\revis{read off from Fig.\ref{fig:2xA13}(a)} is that aggregation
and collapse are not separate processes (near $T\approx 0.97$), 
similar to the corresponding heteropolymer system~\cite{jbj1,jbj2}.

At about
$T\approx 0.24$ (close to the single-chain freezing temperature),
the multiple-chain homopolymer complex crystallizes in a separate process.
This is in strong contrast to the heteropolymer systems, where aggregation, collapse,
{\em and\/} crystallization (hydrophobic-core formation) is a
 single-step process~\cite{jbj1,jbj2}.
In Fig.~\ref{fig:2xA13}(a), we have also included a comparison with a two-chain system
with much stronger bending rigidity $\kappa=10$. As expected, there is
a single aggregation transition near the temperature, where the system of less stiff
semiflexible polymers with $\kappa=0.25$ collapses.
However, a further crystallization process at lower temperatures does not occur:
There is no globular (liquid) pseudophase of defragmented \revis{relatively stiff} 
semiflexible polymers.

\revis{Of course, in finite and small polymer systems conformational macrostates
form no ``phases'' in a strict thermodynamic sense. We hence call the stable macrostates 
in the following ``pseudophases'' in order to emphasize the difference.} 
Nonetheless, 
by introducing a ``phase'' separation parameter
\begin{equation}
\label{eq:gamma}
\Gamma^2=\frac{1}{2M^2}\sum_{\mu,\nu=1}^{M}
\left(\textbf{r}_{\text{COM}}^{(\mu)}-\textbf{r}_{\text{COM}}^{(\nu)}\right)^2\, ,
\end{equation}
\revis{where
$\textbf{r}_{\text{COM}}^{(\mu)}=\sum_{i_\mu=1}^N\textbf{r}_{i_\mu}/N$~\cite{jbj2}
is the center of mass of the polymers}, 
a clear discrimination of the different pseudophases can be made.
Small values \revis{of $\Gamma$} correspond to aggregated and higher values to 
fragmented conformations.
In Fig.~\ref{fig:2xA13}(b), the canonical expectation value $\langle\Gamma\rangle$ and
its fluctuation $d\langle \Gamma\rangle/dT$ are plotted.  
The peak position of $d\langle \Gamma\rangle/dT$ coincides nicely with the corresponding
\revis{peak temperature of the specific heat and thus signals the aggregation transition.}  

\begin{figure}
\onefigure[width=0.49\textwidth]{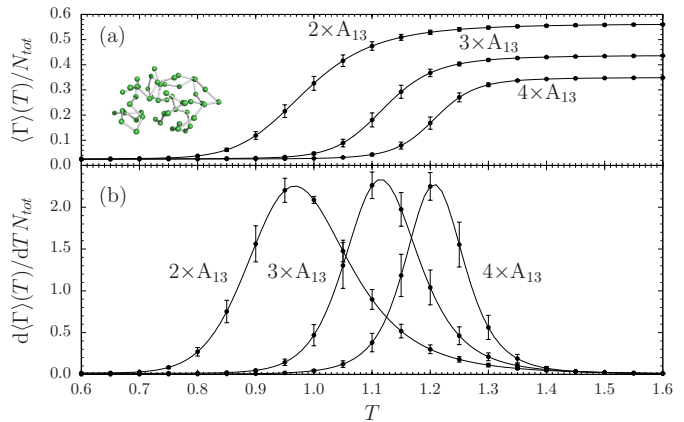}
\caption{\label{fig:gamma}
(a) Canonical expectation values and (b) fluctuations of the aggregation parameter $\Gamma$ for 
multi-chain systems with $\kappa=0.25$. In (a), also an exemplified globular $4\times A_{13}$ aggregate is
depicted.
}
\end{figure}

In Fig.~\ref{fig:gamma}, the aggregation parameter and its fluctuations are 
shown for different system sizes of up to four chains. The peak shifts towards
higher temperatures and gets sharper with increasing number of chains, since
intra-chain and inter-chain monomer-monomer contacts are not \revis{energetically}
distinguished. 
The hypothetic 
maximal total number of intrinsic
contacts is $n_\text{intra}^\text{max}=M(N-2)(N-1)/2\sim MN^2=N N_\text{tot}$, and for the maximally possible 
number of inter-chain contacts one has
$n_\text{inter}^\text{max}=M(M-1)N^2/2\sim M^2N^2=N_\text{tot}^2$~\cite{rem1}. 
For large $M$, the relative fraction $r_\text{inter}$ 
of the inter-chain contacts,
\begin{eqnarray}
\label{eq:ratio}
\hspace*{-5mm}r_\text{inter}&=&
\frac{n_\text{inter}^\text{max}}{n_\text{intra}^\text{max}+n_\text{inter}^\text{max}} \nonumber \\
&=&1-\frac{1}{M}\left(1-\frac{1}{N}\right)\left(1-\frac{2}{N}\right)+{\cal O}\left(\frac{1}{M^2N}\right),
\end{eqnarray}
behaves like $r_\text{inter}\sim 1-M^{-1}$, i.e.,
the relative influence of inter-chain contacts increases rapidly \revis{towards unity} 
with the number of chains. 
In consequence, 
aggregation dominates over collapse of the individual chains. Even for the two-chain system
$2\times A_{13}$ this estimate is reasonable: The energy of the lowest-energy conformation we found
\revis{numerically}
is $E^{(\text{min})}\approx -83.61$ and the contribution of the inter-chain contacts is 
$E_{\text{inter}}^{(\text{min})}\approx -50.20\approx r_\text{inter}^{(\text{min})}\, E^{(\text{min})}$
with $r_\text{inter}^{(\text{min})}\approx 0.60$. This 
coincides nicely with the corresponding value of the above contact ratio,
$r_\text{inter}\approx 0.56$. In fact, considering energetic and
structural fluctuations of the larger systems \revis{with three and four chains}, 
we have not found indications for an additional 
collapse transition at temperatures higher than the aggregation transition.

\begin{figure}
\onefigure[width=0.49\textwidth]{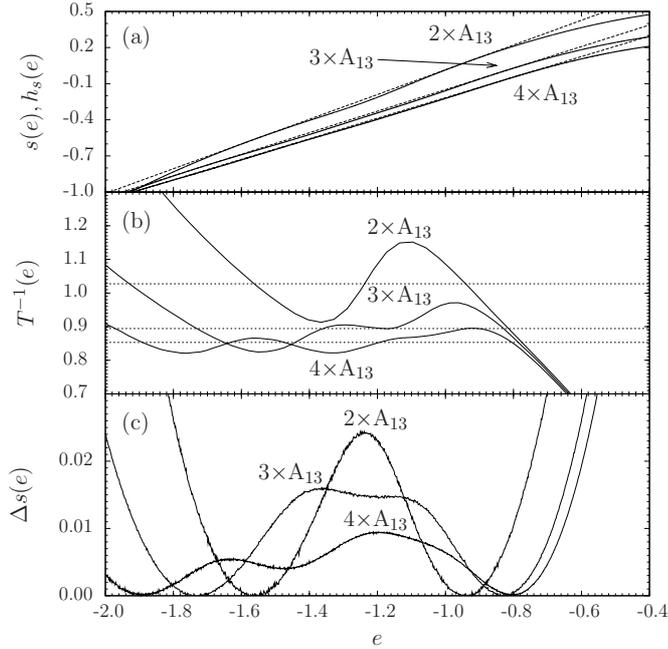}
\caption{\label{fig:gE}
(a) Microcanonical entropy per monomer $s(e)$ and the Gibbs constructions $h_s(E)$ (dashed lines), 
(b) reciprocal
caloric temperatures and Maxwell constructions, and 
(c) relative surface entropies per monomer $\Delta s_\text{surf}$.}
\end{figure}
Below the aggregation temperature, the entropic loss of the individual chains is overcompensated
by the energetic gain of forming a joint globular aggregate. However, the entropic change
while passing the aggregation transition is noticeably smaller than what we found recently for
heteropolymer systems, where no separate freezing transition occurs~\cite{jbj2}. In the intermediate
fluid ``globular'' pseudophase, the aggregate \revis{of the homopolymers} 
thus behaves like a single chain of length $MN$. 
Consequently, reducing the temperature, the aggregate optimizes the monomer arrangements
in order to maximize energetic contacts. \revis{Indicated by the peak in the specific heat near 
$T\approx 0.2$ (see  
Fig.~\ref{fig:2xA13} and also Ref.~\mbox{\cite{baj1}})}, the small globular aggregates
freeze into spherical amorphous structures with a maximum number of 
inter-chain contacts. 
For rather stiff
semiflexible polymers \revis{(as our example with $\kappa= 10$)}, 
however, a separate freezing transition does not occur 
\revis{and the peak of $c_V$ near $T\approx 0.9$ indicates a single-step 
transition from rod-like coils to a liquid-crystal-like phase}~\cite{stukanA}.

In our recent studies of heteropolymer aggregation~\cite{jbj1,jbj2}, we also observed a 
pronounced single transition \revis{of a different nature. In this case,} 
the formation of a heteropolymer complex 
consisting of different chains with compact hydrophobic core
is roughly a single-step process, because hydrophobic-core
formation (``freezing'') favors conformations with very small entropy. 
For semiflexible homopolymer 
systems, we find in the rather floppy limit that the \emph{freezing} temperature 
($T\approx 0.2$, for single chains or globular aggregates with $\kappa = 0.25$) is almost identical 
with the \emph{aggregation} temperature for heteropolymer systems~\cite{jbj1,jbj2}. 
\revis{This coincidence in the behavior of these different systems is due to 
the formation of a single, very compact hydrophobic domain (the monomers of the 
interacting homopolymers are as hydophobic as the A-type monomers of the heteropolymer) 
which maximizes the number of energetic contacts. Stiff homopolymers, on the
other hand,  cannot form a maximally compact hydrophobic core and hence do not 
crystallize in a separate transition.}

\begin{table*}
\begin{center}
\begin{tabular}{ccccccc}
\hline\hline
System & $T_\text{agg}$ & $\Delta s_\text{surf}$ &
$e_\text{agg}$ & $e_\text{frag}$ & $\Delta q$ & $\Delta q/T_\text{agg}$\\ 
\hline
2$\times$A$_{13}$ & $0.973$  & $0.024$ & $-1.566$ & $-0.944$ & $0.622$ & $0.639$\\
3$\times$A$_{13}$ & $1.118$  & $0.016$ & $-1.730$ & $-0.831$ & $0.899$ & $0.804$\\
4$\times$A$_{13}$ & $1.172$  & $0.009$ & $-1.892$ & $-0.799$ & $1.093$ & $0.932$\\
\hline\hline
\end{tabular}
\end{center}
\caption{\label{tab:lsys} Aggregation temperatures $T_\text{agg}$, 
relative surface entropies per monomer $\Delta s_\text{surf}$,
relative aggregation and fragmentation energies per monomer, $e_\text{agg}$ and $e_\text{frag}$,
respectively, latent heat per monomer $\Delta q$, and phase-separation entropy per monomer
$\Delta q/T_\text{agg}$.}
\end{table*}
For a deeper understanding of the aggregation transition,
we now analyze entropic effects accompanying this 
transition in the microcanonical ensemble for an isolated system \revis{of multiple semiflexible 
chains with $\kappa = 0.25$.}
The microcanonical entropies per monomer, $s=S/N_\text{tot}$,
are shown in Fig.~\ref{fig:gE}(a) as functions of the total
energy per monomer, $e=E/N_\text{tot}$. The 
reciprocal caloric temperature, $T^{-1}(E)=\partial S/\partial E$, is plotted 
in Fig.~\ref{fig:gE}(b). The plots reveal an exciting phenomenon: Increasing the energy
entails a reduction of temperature \revis{in the transition region, known as the 
backbending effect}~\cite{gross1}. \revis{This signals a phase-separation process~\cite{jbj1,jbj2}
which is caused by surface effects reducing the}
entropy which, in an isolated system, results in a decrease of temperature by increasing the total system
energy~\cite{gross1,thirring1,wales1,janke1,hilbert1,pleimling1}.
This phenomenon is called ``backbending effect'', because the caloric temperature curve
changes in the transition region its monotonic behavior with increasing total energy~\cite{gross1}.
For truly isolated systems, this is a physical effect which
has been verified, e.g., in atomic clustering experiments~\cite{schmidt1}.

In our previous studies of heteropolymer aggregates~\cite{jbj1,jbj2}, we took technical 
advantage of 
this
microcanonical view 
to identify
the peptide aggregation
as so-called ``coupled binding-folding transition'', i.e., the individual heteropolymer
chains refold \revis{during the binding process}
and the finally formed aggregate possesses a single
compact hydrophobic domain.

By closer inspection of Fig.~\ref{fig:gE}(b), we find 
for the homoploymer system a hiercharchical substructure
caused by these surface effects. The \revis{number of oscillations} of the curves, 
increasing with system size, reveals that the aggregation transition is actually a 
composition of different subprocesses, each of which
is an individual phase-separation process. \revis{The amplitude of these 
oscillations decreases with system size showing that these subprocesses comprise 
a smaller surface-entropic barrier [see Fig.~\ref{fig:gE}(c)].}

The $2\times A_{13}$ system exhibits a single backbending 
effect as only two
chains aggregate. For the three-chain system $3\times A_{13}$, a different scenario is apparent.
In the higher-energy regime, first two chains stick together, and the formation of the 
three-chain globule is a separate process at lower energies. This hierarchical procedure 
continues for larger systems as, e.g., for $4\times A_{13}$. However, the impact of the 
individual backbending effects is getting weaker and, in the thermodynamic limit, 
these effects are expected to disappear asymptotically, whereas the first-order character
of this transition remains. The horizontal lines in Fig.~\ref{fig:gE}(b) 
are the respective Maxwell constructions
and define the aggregation temperatures $T_\text{agg}$. In the following, we denote 
the leftmost (rightmost) energy where $T^{-1}(e)=T^{-1}_\text{agg}$ as $e_\text{agg}$ 
($e_\text{frag}$). In the entropy curves, the Maxwell constructions correspond to concave hulls
$h_s(e)$ with $\partial h_s(e)/\partial e=T^{-1}_\text{agg}$ (Gibbs constructions) in the transition regime 
[see Fig.~\ref{fig:gE}(a)], where entropy is reduced due to surface effects
(\revis{the convex region of the microcanonical entropy curve is sometimes called} ``convex intruder''~\cite{gross1}). 

For a quantitative analysis, we define 
$\Delta s(e)=h_s(e)-s(e)$ which is plotted in Fig.~\ref{fig:gE}(c). Within the 
transition region (i.e., for $e_\text{agg}\le e \le e_\text{frag}$), the
peak height 
$\Delta s_\text{surf}=\max_{e_\text{agg}\le e \le e_\text{frag}} \Delta s(e)$ defines the surface entropy. 
The energetic
width of the phase-coexistence regime is the latent heat per monomer, 
$\Delta q=e_\text{frag}-e_\text{agg}=T_\text{agg}[s(e_\text{frag})-s(e_\text{agg})])$.
Thus the entropic \revis{phase separation} barrier $\Delta q/T_\text{agg}$ should survive in the thermodynamic
limit, if the aggregation of semiflexible polymers is a first-order-like phase-separation process
with coexistence of aggregates and fragments.

Values for these quantities are listed in Table~\ref{tab:lsys} for the three polymer systems
considered in our study. We find that with increasing
system size the surface entropies $\Delta s_\text{surf}$ decrease and thus
the \revis{influence of surface effects} is getting weaker. On the other hand, the latent heat per monomer
increases, supporting 
the first-order character of
the aggregation transition.

We have shown in this Letter that
the aggregation of interacting semiflexible polymers is a first-order phase-separation process. 
\revis{
For two reasons, we focused on small systems with
up to four chains with 13 monomers
in a cubic box of edge length
$L=60$. The first reason is that one of the primary goals of this study was
the unraveling of underlying mechanisms in multiple-chain aggregation processes. This also
includes the identification of sequentially ordered, i.e., hierarchical, subphase transitions
accompanying the phase separation process of aggregation. As our
microcanonical analysis revealed, these transitions exhibit peculiarities compared to
thermodynamic phase transitions, which are due to dominant entropic surface effects in the small
systems considered here. The surface changes at the interface of co-existing
aggregates and fragments let the microcanonical temperature decrease while energy increases,
whereas far away from the transition region the more intuitive behavior is observed: temperature
increases with energy. Because of the monotonic change of the microcanonical temperature
in the transition region, this phenomenon is called ``backbending''.
A second reason for restricting ourselves to small systems is the demand
for a particularly high accuracy needed for the precise estimation of the density of states
which is the basis for the microcanonical analysis.}
The \revis{observed} surface effects are expected to
vanish in the thermodynamic limit~\cite{jbj2}. However, for molecular small-scale
applications based on the interplay of only a few molecules
and the understanding of biomolecular aggregation processes, the
thermodynamics of finite-size effects is relevant.

Our precise microcanonical analysis revealed that \revis{for homopolymers} the aggregation 
transition is accompanied by hierarchical
backbending effects. We found that after aggregation, complexes of rather floppy semiflexible polymers 
behave like globules in the liquid regime and freeze at low temperatures in a separate process.
In contrast, for rather stiff homopolymers, no separate freezing transition is observed, in coincidence
with former results~\cite{stukanA}.
\revis{The overall transition behavior of semiflexible homopolymers is also different compared to the
previously studied heteropolymer model~\cite{jbj1,jbj2} which only differs in the sequential disorder
of hydrophobic and polar monomers, whereas in the homopolymer case all monomers are treated as hydrophobic.
However, we found similarities in the transition towards the formation of a single hydrophobic domain,
provided the stiffness of the homopolymer chain is sufficiently weak.}

The understanding of aggregation and crystallization processes of polymers is a necessary 
prerequisite for the design of technological applications in material science as, e.g., 
ordered nanoscopic structures like fibers, nanopores, and channels or
amorphous polymeric cells with particular electronic properties. 
Aggregation processes are also essential in biological systems, where enzymatic and 
motoric action is mediated by molecular binding processes. Furthermore, molecular cluster
formation can also cause disastrous diseases like Alzheimer's, which is yet another reason, why 
generic features of polymer-polymer interactions are worth
being studied.

This work is partially supported by the DFG (German Science Foundation) under Grant
Nos.\ JA \mbox{483/24-1/2/3}, the Leipzig Graduate School of Excellence ``BuildMoNa'', 
and by the Max Planck Society.
Support by a NIC supercomputer time grant (No.~hlz11) of the Forschungszentrum J{\"u}lich 
is acknowledged. 


\begin{thebibliography}{99}
\bibitem{rem1}
Due to excluded volume and optimal space-filling constraints, these numbers need 
to be scaled by about $q/N_\text{tot}$, where $q$ is the effective coordination
 number (which is, e.g., $q\approx 10.0$ for a perfect fcc crystal).
\end{thebibliography}
\end{document}